\documentclass[%
 reprint,
superscriptaddress,
 amsmath,amssymb,
 aps,
floatfix,
showkeys
]{revtex4-2}

\usepackage{graphicx}% Include figure files
\usepackage{dcolumn}% Align table columns on decimal point
\usepackage{bm}% bold math
\begin{document}

\preprint{APS/123-QED}

\title{AI-Assisted Topological High-Harmonic Spectroscopy for\\ Characterizing Crystal Symmetries and Nonlinear Responses}% Force line breaks with \\

\author{David Hervalejo-Sancho}
\email[E-mail: ]{davidhersan@usal.es}
\affiliation{Grupo de Materiales Nanoestructurados para Aplicaciones en Optoelectrónica, Departamento de Física Aplicada, Universidad de Salamanca, Salamanca E-37008, Spain.
}
\affiliation{Unidad de Excelencia en Luz y Materia Estructuradas (LUMES), Universidad de Salamanca, Salamanca E-37008, Spain.}

\author{Sergio Mart\'in-Domene}

\affiliation{Unidad de Excelencia en Luz y Materia Estructuradas (LUMES), Universidad de Salamanca, Salamanca E-37008, Spain.}
\affiliation{Grupo de Investigación en Aplicaciones del Láser y Fotónica, Departamento de Física Aplicada, Universidad de Salamanca, Salamanca E-37008, Spain.
}

\author{Rodrigo Mart\'in-Hern\'andez}%

\affiliation{Unidad de Excelencia en Luz y Materia Estructuradas (LUMES), Universidad de Salamanca, Salamanca E-37008, Spain.}
\affiliation{Grupo de Investigación en Aplicaciones del Láser y Fotónica, Departamento de Física Aplicada, Universidad de Salamanca, Salamanca E-37008, Spain.
}
\author{Antonio Pic\'on}%
\affiliation{Instituto de Ciencia de Materiales de Madrid (ICMM-CSIC), E-28049 Madrid, Spain
}

\author{Luis Plaja}
\affiliation{Unidad de Excelencia en Luz y Materia Estructuradas (LUMES), Universidad de Salamanca, Salamanca E-37008, Spain.}
\affiliation{Grupo de Investigación en Aplicaciones del Láser y Fotónica, Departamento de Física Aplicada, Universidad de Salamanca, Salamanca E-37008, Spain.
}

\author{Jos\'e M. Caridad}%
\affiliation{Grupo de Materiales Nanoestructurados para Aplicaciones en Optoelectrónica, Departamento de Física Aplicada, Universidad de Salamanca, Salamanca E-37008, Spain.
}
\affiliation{Unidad de Excelencia en Luz y Materia Estructuradas (LUMES), Universidad de Salamanca, Salamanca E-37008, Spain.}

\author{Javier Serrano}%

\affiliation{Unidad de Excelencia en Luz y Materia Estructuradas (LUMES), Universidad de Salamanca, Salamanca E-37008, Spain.}
\affiliation{Grupo de Investigación en Aplicaciones del Láser y Fotónica, Departamento de Física Aplicada, Universidad de Salamanca, Salamanca E-37008, Spain.
}
\author{Carlos Hern\'andez-Garc\'ia}
\affiliation{Unidad de Excelencia en Luz y Materia Estructuradas (LUMES), Universidad de Salamanca, Salamanca E-37008, Spain.}
\affiliation{Grupo de Investigación en Aplicaciones del Láser y Fotónica, Departamento de Física Aplicada, Universidad de Salamanca, Salamanca E-37008, Spain.
}

\date{\today}% It is always \today, today,

\begin{abstract}
Structured light has emerged as a unique tool for probing nonlinear light–matter interactions. Particularly interesting is its interaction with anisotropic media, such as crystalline materials, whose anisotropic responses encode information about their electronic properties. Recently, topological high-harmonic spectroscopy was introduced as an all-optical approach to extract the anisotropic nonlinear response of graphene from harmonic spatial intensity profiles.
Here, topological high-harmonic spectroscopy is combined with artificial intelligence (AI) to retrieve the anisotropic nonlinear response and crystal symmetry of atomically thin materials. A convolutional neural network is trained and validated to learn the inverse mapping between harmonic far-field intensity profiles and the nonlinear response of materials exhibiting different rotational symmetries. Comparisons with microscopic high-harmonic-generation simulations demonstrate the accurate retrieval of distinct effective nonlinear responses in single-symmetry crystals, such as single-layer graphene and bilayer graphene with different stacking orders, and in crystals exhibiting multiple symmetry components, such as monolayer hexagonal boron nitride. Macroscopic high-harmonic-generation simulations assess the robustness and applicability limits of the technique. These results establish AI-assisted topological high-harmonic spectroscopy as a fast, contactless tool for characterizing nonlinear optical responses and crystal symmetries without phase-resolved measurements, opening prospects for applications to complex systems such as polycrystals and van der Waals heterostructures. 

\end{abstract}

\keywords{structured light, nonlinear optics, two-dimensional materials, high-harmonic generation, harmonic spectroscopy, light topology, artificial intelligence}
\maketitle

\section{Introduction} 

Nonlinear optics have undergone great development since the first observations of \textit{high harmonic generation} (HHG) in gases, more than 30 years ago \cite{mcphersonStudiesMultiphotonProduction1987,ferrayMultipleharmonicConversion10641988}. This extreme nonlinear phenomenon has made it possible to obtain high-frequency coherent radiation that can extend into the soft X-ray range \cite{popmintchevBrightCoherentUltrahigh2012}, with well-defined phase relationships that enable its synthesis into light pulses in the attosecond domain \cite{paulObservationTrainAttosecond2001, hentschelAttosecondMetrology2001,varillas2026roadmapattosecondscience}. In addition to generating ultrashort laser pulses, one of the most notable applications of HHG is high-harmonic spectroscopy, which allows probing the ultrafast dynamics of a wide variety of systems on previously inaccessible attosecond timescales \cite{krauszAttosecondPhysics2009, itataniTomographicImagingMolecular2004, silvaHighHarmonicImaging2018}. The extension of HHG to solid-state systems \cite{ghimireObservationHighorderHarmonic2011} has significantly expanded the scope of high-harmonic spectroscopy, opening new avenues for investigating the attosecond electron dynamics that govern the physical properties of solids.

Solid-state HHG can be understood by adapting the \textit{three-step model} \cite{schaferThresholdIonizationHigh1993, corkumPlasmaPerspectiveStrong1993} from gases to solids in reciprocal space \cite{yueIntroductionTheoryHighharmonic2022}. In the first step, the intense driving laser field generates an electron-hole pair through tunnel excitation (adiabatic process) or via direct crossing through degenerated energy points in gapless materials (non-adiabatic). In the second step, the electron and hole are accelerated by the laser field in momentum space within their respective energy bands. In the final step, recombination of the electron-hole pair leads to the emission of high-order harmonics. In addition, the nonlinear intraband currents arising from carrier motion within nonparabolic bands give rise to additional harmonic contributions \cite{goldeHighHarmonicsGenerated2008}. As in gases, the nonlinear nature of HHG makes it highly sensitive to both material and laser field properties. However, more intricate phenomena arise in solids. For example, lattice periodicity and a more complex electronic band structure, compared with the discrete energy levels of atoms, give rise to features such as different scaling law for the cutoff frequency---the maximum harmonic photon energy emitted---or multiple non-perturbative plateaus \cite{ghimireHighharmonicGenerationSolids2019}. In addition, lattice anisotropy leads to a strong dependence of harmonic emission on the polarization state of the driving field. This effect was first reported in bulk ZnO \cite{ghimireObservationHighorderHarmonic2011}, where the non-centrosymmetric wurtzite structure generates, overall, both odd and even harmonics, but only odd harmonics along the optical axis. Originating from the complex electron dynamics driven by the periodic lattice potential, this phenomenon has been investigated both experimentally \cite{Schubert2016, You2017} and theoretically \cite{Tancogne2017}, demonstrating its potential to reveal information about the electronic band structure and interatomic bonding of the target material. Another readily accessible property is lattice symmetry, which can be probed through HHG in materials such as GaAs \cite{kaassamaniPolarizationSpectroscopyHighOrder2022} or \textit{single-layer graphene} (SLG) \cite{zurron-cifuentesOpticalAnisotropyNonperturbative2019}, each exhibiting periodic dependence with polarization consistent with their four-fold and six-fold symmetries, respectively.

The anisotropic behaviour of HHG in solids makes the use of structured driving beams \cite{bliokh2023, shen2023} particularly interesting. Structured light refers to the spatial tailoring of the amplitude, phase, and polarization distributions of light beams, allowing control over their angular momentum. We distinguish between (1) \textit{spin angular momentum} (SAM), related to the polarization state of light and taking values $\hbar \sigma$ per photon, with $\sigma=-1$ for \textit{right-circular polarization} (RCP) and $\sigma=1$ for \textit{left-circular polarization} (LCP); and (2) \textit{orbital angular momentum} (OAM), associated to the spatial phase distribution of a beam. For beams with an azimuthal phase dependence $\exp(i\ell\phi)$---also known as vortex beams---the associated OAM is $\ell\hbar$ per photon, where the \textit{topological charge} $\ell$ denotes the number of $2\pi$ phase windings along the azimuthal coordinate $\phi$ around the beam axis. OAM conservation in HHG in gases \cite{hernandez-garciaAttosecondExtremeUltraviolet2013a, gariepyCreatingHighHarmonicBeams2014, Geneaux2016, martinhernandez2025, Porras2026} and solids \cite{gauthierOrbitalAngularMomentum2019} enables the generation of harmonic beams with rich OAM content.

However, among the various structured beam configurations, \textit{linearly-polarized vector beams} (LPVB) \cite{zhanCylindricalVectorBeams2009} are particularly interesting because of their complex nonlinear interaction with anisotropic materials. 
LPVBs can be decomposed into two RCP and LCP vortex beams with opposite $\ell$, therefore resulting in zero net OAM. They can be further characterized by the \textit{Poincaré index} ($\mathcal{P}$) \cite{guntherIntroductionVectorField2021}, which takes integer values corresponding to the number of complete rotations of the polarization direction along the azimuthal coordinate, with $|\mathcal{P}|=|\ell|$. 
The topological properties of LPVB---such as their Poincaré index and the topological charges of their circularly polarized components---are preserved in HHG in gases \cite{hernandez-garciaExtremeUltravioletVector2017}, since the local response of an isotropic gas is independent of the orientation of the driving field's linear polarization. In solid targets, however, the lattice-dependent anisotropic response involves a much richer phenomenology.

In a recent work \cite{garcia-cabreraTopologicalHighharmonicSpectroscopy2024}, it was theoretically demonstrated that the topology of high-order harmonics driven by LPVB in SLG---an archetypal two-dimensional material composed of carbon atoms arranged in a honeycomb lattice---differs substantially from that of the driving beam. In particular, the harmonic intensity profiles exhibit a six-fold symmetry, reflected in the appearance of additional OAM contributions in their RCP and LCP components, which are directly related to the anisotropic non-linear response of the SLG target. Moreover, the interband and intraband contributions were found to exhibit distinct harmonic topologies, revealing the sensitivity of harmonic topology to subcycle electronic dynamics. These findings motivated the proposal of \textit{topological high-harmonic spectroscopy}, a technique to probe the ultrafast anisotropic properties of solid materials through far-field intensity measurements of the RCP/LCP harmonic beams. Despite the potential of this technique for material characterization, an analytical solution to the inverse problem---retrieving material properties from the far-field intensity profiles---was demonstrated only for the specific case of SLG, limiting its applicability.

In this work, we integrate the use of artificial intelligence (AI) into topological high-harmonic spectroscopy as a universal framework for solving the inverse problem and quantitatively retrieving the anisotropic nonlinear response of atomically-thin crystalline materials. Neural networks have already been applied in nonlinear and ultrafast optics \cite{whiteAttosecondStreakingPhase2019, lytovaDeepLearningHigh2023, zahavyDeepLearningReconstruction2018, pablos-marinSimulatingMacroscopicHighorder2023}, making them a promising approach for accessing information that would otherwise be difficult to explore. Here, we design, train and validate a convolutional neural network using purpose-built datasets to retrieve quantitative information about the non-linear response of crystal lattices with arbitrary symmetry from the far-field intensity profiles of the harmonic beams. We use it to reconstruct the nonlinear response of materials exhibiting twofold, threefold, fourfold, or sixfold rotational symmetries, thus including the nontrivial rotational symmetries allowed in two-dimensional periodic crystals \cite{ubicCrystallographyCrystalChemistry2024}. We validate the network's predictions in two complementary ways. First, we compare them with microscopic, fully quantum simulations of solid targets using the \textit{semiconductor Bloch equations} (SBE). We consider systems with the same sixfold rotational symmetry but different nonlinear responses---SLG and bilayer graphene (BLG) with different stacking orders---as well as systems as monolayer hexagonal boron nitride (hBN), which exhibits multiple rotational symmetries---threefold and sixfold in this case. Second, we perform macroscopic HHG simulations in SLG using the electromagnetic field propagator, accounting for harmonic propagation and phase-matching to assess the applicability of the technique under experimentally relevant conditions. 

Our results demonstrate that AI-assisted topological high-harmonic spectroscopy can retrieve the lattice symmetry and the anisotropic nonlinear response of crystalline materials. These findings may prove relevant to study materials exhibiting multiple coexisting symmetries, such as van der Waals heterostructures or polycrystals, to probe exciton dynamics, or to detect lattice defects.

\section{Methodology} 

\begin{figure*}
\centering
\includegraphics[width=0.99\textwidth]{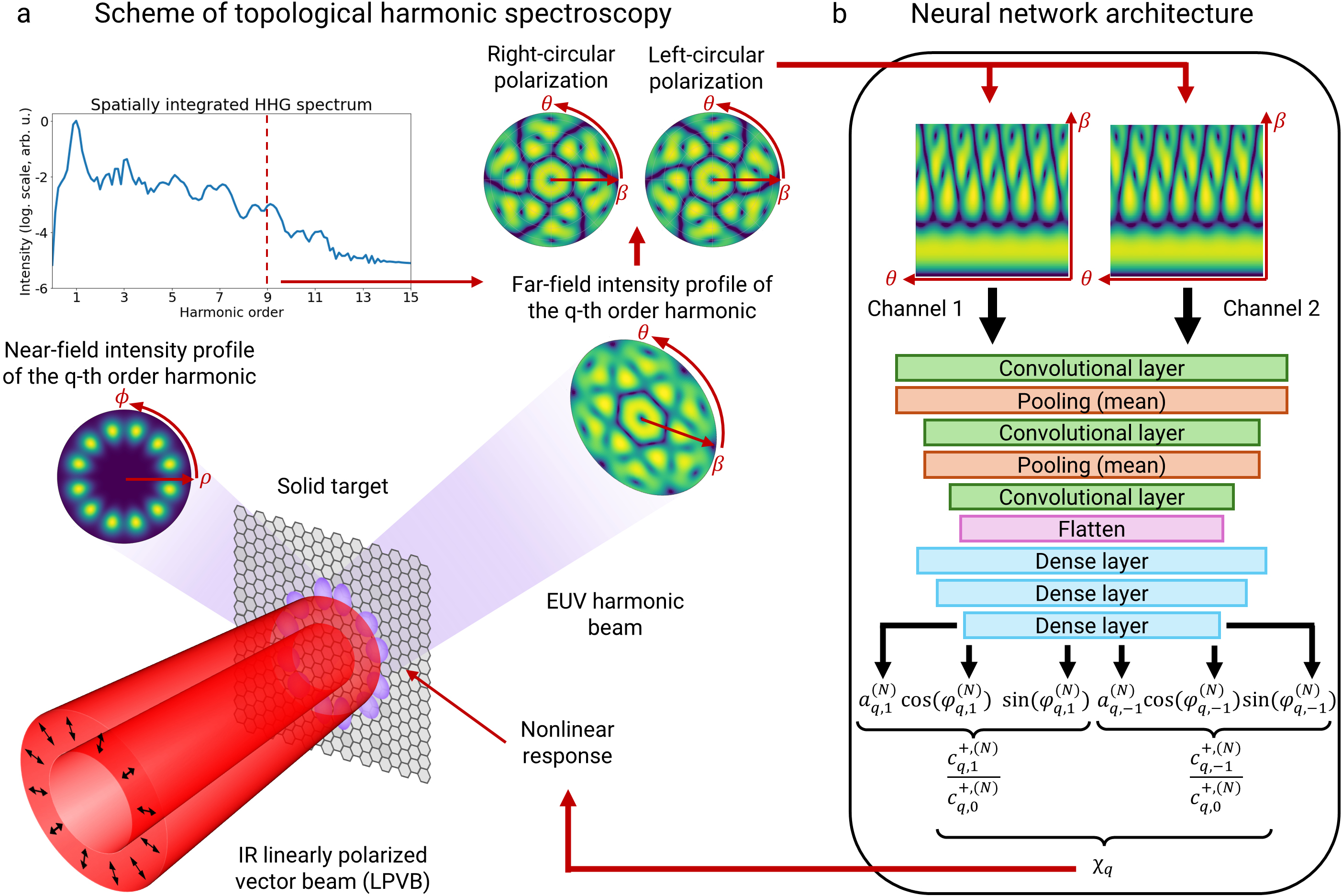}
\caption{Sketch of AI-assisted topological high-harmonic spectroscopy. a) An intense infrared LPVB with Poincaré index $\mathcal{P}$ is focused on a thin crystal layer, generating high-order harmonics through its strongly nonlinear interaction with the material. The upper-left inset shows an example of the spatially integrated HHG spectrum generated in SLG driven by an LPVB with a wavelength of $3\,\mu\mathrm{m}$ and a peak intensity of $5\times10^{10}\,\mathrm{W\,cm^{-2}}$. The anisotropic crystal response is imprinted on the harmonic near field, described by coordinates $\rho$ and $\phi$. After propagation to the far field, described by coordinates $\beta$ and $\theta$, the spatial intensity profile of each harmonic is recorded, where no phase information is required. The RCP and LCP intensity profiles of a selected $q$th-order harmonic serve as inputs to the neural network shown in panel b). The neural network is composed of convolutional and dense layers and is trained to retrieve the nonlinear response associated with each crystal symmetry $N$. For each $N$, six output neurons encode the amplitudes and phases of the two relative Fourier coefficients $c^{\pm,(N)}_{q,s}/c^{\pm,(N)}_{q,0}$, with $s= \{ -1,1 \}$, from which the effective nonlinear susceptibility associated with the $q$th harmonic is predicted.}
\label{fig:Setup}
\end{figure*}

The concept of AI-assisted topological high-harmonic spectroscopy is illustrated in Fig. \ref{fig:Setup}a. A mid-infrared femtosecond LPVB drives HHG in a solid-state system (as an example, $\mathcal{P}=1$ in the sketch), generating high-order harmonics. The near-field intensity profile of each harmonic encodes information of the anisotropic nonlinear response. The harmonics propagate to a far-field detector, where their spatial intensity profiles are recorded. The top-left inset shows an example of the spatially integrated harmonic spectrum obtained in SLG, extending up to the ninth order. The far-field spatially-resolved intensity profile of a selected $q$-th order harmonic is decomposed into its RCP and LCP components. The spatial RCP and LCP intensity profiles are then fed into a convolutional neural network (Fig. \ref{fig:Setup}b), which predicts the near-field nonlinear response of the solid target. This response encodes information about the lattice symmetry and nonlinear anisotropy of the crystal, enabling its characterization. In this section, we describe how the nonlinear response is retrieved, the architecture of the neural network, and the HHG models used to validate its predictions.

\subsection{Retrieving the nonlinear response from topological high-harmonic spectroscopy}
\label{sec:FarFieldModel}

Here we describe how topological high-harmonic spectroscopy enables the retrieval of the nonlinear response of a solid target, following the framework presented in \cite{garcia-cabreraTopologicalHighharmonicSpectroscopy2024}. We restrict our analysis to thin crystals, neglecting propagation effects within the target. We consider a LPVB with Poincaré index $\mathcal{P}$ driving HHG in a layer with $N$-fold lattice symmetry.

The RCP/LCP components of the harmonics contain OAM modes directly related to the lattice symmetry $N$ \cite{garcia-cabreraTopologicalHighharmonicSpectroscopy2024}. Therefore, the near-field RCP and LCP amplitudes of the $q$-th order harmonic at the target, $F_q^\pm(\rho, \phi)$, where  $\rho$ and $\phi$ are the radial and azimuthal near-field coordinates, respectively, can be expressed as a superposition of OAM modes weighted by complex Fourier coefficients, $c_{q,s}^{\pm}$, as
\begin{equation}
	F_q^\pm(\rho, \phi)=e^{\pm i\mathcal{P}\phi}\sum_{s=-1}^1 c_{q,s}^{\pm}(\rho)e^{\pm iNs\mathcal{P}\phi},
	\label{eq:nearf}
\end{equation}
where the upper and lower signs correspond to the RCP and LCP components, respectively. From the azimuthal phase dependence in Eq. \ref{eq:nearf}, each $s$ term carries an OAM characterized by the topological charge $\ell^s=\pm(1+Ns)\mathcal{P}$. Thus, the OAM of the $s=0$ term is inherited from the LPVB driving beam, whereas the $s\in \{-1,1 \}$ terms represent additional OAM components that encode the lattice symmetry and are absent in the driving beam. The complex Fourier coefficients are directly related to an effective nonlinear susceptibility through

\begin{equation}
	\chi_q^\pm(\rho, \phi)\propto\sum_{s=-1}^1 c_{q,s}^{\pm}(\rho)e^{\pm iNs\mathcal{P}\phi}.
	\label{eq:sus}
\end{equation}

We obtain the far-field harmonic profiles at the detector through the Fraunhofer diffraction integral. The expression can be simplified assuming that high-order harmonics are generated on the ring of maximum intensity of the driving LPVB, at $\rho=\rho_0$, an approximation previously used in experimental studies of HHG driven by structured light beams \cite{regoGenerationExtremeultravioletBeams2019, regoNecklacestructuredHighharmonicGeneration2022, martin-hernandezEnhancedFocusingAttosecond2026}. Under this approximation, the RCP and LCP amplitudes of the $q$th-order harmonic, $U_q^{\pm}(\beta,\theta)$, where $\beta$ and $\theta$ are the far-field divergence and azimuthal coordinates, are given by 
\begin{equation}
    \begin{split}
	U_q^{\pm}(\beta,\theta)=&\frac{\kappa}{i2\pi}\frac{e^{i\kappa D}}{D}e^{i\kappa \beta^2 D/2}\\
          &\times\int F_q^\pm(\rho, \phi)\delta(\rho-\rho_0)e^{-i\kappa \rho \beta cos(\theta-\phi)}\rho \mathrm{d}\rho \mathrm{d}\phi, %d\rho d\phi,
	\label{eq:Fran}
    \end{split}
\end{equation}

where $\kappa=2\pi q/\lambda$ is the harmonic wave number, $\lambda$ being the wavelength of the driving beam, and $D$ the distance from the solid target to the detector. Performing radial integration in Eq. \ref{eq:Fran}, applying the Jacobi-Anger expansion, and neglecting the quadratic phase factor associated with the spherical wavefront, we obtain
\begin{equation}
\begin{split}
	U_q^{\pm}(\beta,\theta)=&-i\kappa\frac{e^{i\kappa D}}{D}\rho_0 \sum_{s=-1}^1 c_{q,s}^{\pm}(\rho_0)e^{\mp i(Ns+1)\mathcal{P}\pi/2} \\
    & \times J_{\pm (Ns+1)\mathcal{P}}\left(\kappa \rho_0 \beta \right)e^{\pm i (Ns+1)\mathcal{P}\theta},
	\label{eq:farf}
\end{split}
\end{equation}
where $J_m$ denotes the Bessel function of the first kind of integer order $m$. Eq. \ref{eq:farf} directly relates the complex Fourier coefficients---and thus, the effective non-linear susceptibility---to the far-field amplitude of the $q$-th order harmonic. Experimental parameters such as $\lambda$, $\rho_0$ or $D$ enter the propagation expression only through an overall complex prefactor or a scaling of the Bessel function argument. For a given set of Fourier coefficients, the normalized far-field intensity patterns therefore retain their shape upon appropriate rescaling of the radial coordinate. Considering non-chiral responses, we can further assume equal coefficients for the LCP and RCP components, $c_{q,s}^{+}=c_{q,s}^{-}$. This formulation allows us to generate datasets based on the complex Fourier coefficients to train a neural network to retrieve information about the solid target from the far-field intensity profiles of its harmonic emission.

\subsection{Neural network architecture}

Our neural network architecture consists of convolutional and dense layers as illustrated in Fig. \ref{fig:Setup}b. Such architectures have previously been used for image-processing tasks \cite{zahavyDeepLearningReconstruction2018, whiteAttosecondStreakingPhase2019}. Under the conditions considered in this work (see below), the best-performing architecture comprises three convolutional layers with 64, 32 and 16 kernels, respectively, followed by two dense layers with 256 and 128 neurons, respectively.

The input consists of two $32\times32$ RGB images representing the normalized RCP and LCP far-field intensity profiles of a given harmonic order $q$, displayed on a logarithmic scale spanning two orders of magnitude. For a given lattice symmetry $N$, the nonlinear response is described by the amplitude and phases of the three Fourier coefficients, $c^{\pm,(N)}_{q,s}$ with $s=\{-1,0,1\}$, as introduced in Eq. \ref{eq:nearf}. Taking the fundamental OAM term $(s=0)$ as reference, we retrieve the two relative coefficients, defined in terms of their amplitudes and phases as
\begin{equation}
\frac{c^{\pm,(N)}_{q,s}}{c^{\pm,(N)}_{q,0}}=a_{q,s}^{(N)}e^{i\varphi_{q,s}^{(N)}},
\qquad s\in \{-1,1\}.
\end{equation}

We represent each phase, $\varphi_{q,s}^{(N)}$, by its sine and cosine to avoid the discontinuity at $0/2\pi$. Accordingly, the output layer for a given lattice symmetry $N$, contains six neurons encoding the amplitude and the sine and cosine of the phase of each relative Fourier coefficient. Our neural network is trained to retrieve the nonlinear response associated with four crystal symmetries, $N=\{2, 3, 4, 6\}$, corresponding to the rotational symmetries allowed in two-dimensional periodic crystals. The output layer thus contains 24 neurons, 6 for each symmetry (see Fig. \ref{fig:Setup}). We select the ninth harmonic, $q=9$, as an example of a non-perturbative harmonic of the crystals studied in this work. 

Dataset generation and neural network training are performed in Python using Keras and TensorFlow on an NVIDIA A30 GPU. Each dataset contains paired inputs and target outputs. We generate the datasets by scanning the Fourier coefficient parameters over ranges determined from HHG simulations (see the following section) and calculating the corresponding inputs---the RCP and LCP far-field harmonic intensity profiles---using Eq. \ref{eq:Fran}. To generate our dataset, we first conduct a parameter grid for each symmetry, using 10 amplitude values spanning $a_{q,s}^{(N)}\in[0,1]$ and 20 phase values spanning $\varphi_{q,s}^{(N)}\in[0,2\pi)$ for each of the two complex coefficients, $s$, while setting the coefficients associated with the other symmetries to zero. This yields 40,000 cases per symmetry, and  160,000 single-symmetry cases in total, each containing a pair of RCP and LCP far-field intensity profiles. Whenever a coefficient has zero amplitude, its phase is also set to zero to avoid ambiguity. The $32\times32$ input intensity profile spans $\beta\in[0,27.7]$ $\mathrm{mrad}$ and $\theta\in[0,2\pi)$ in the radial and azimuthal far-field coordinates, respectively. Note that the upper limit of $27.7$ mrad  corresponds to the first zero of the highest order Bessel function in Eq. \ref{eq:farf}.

We then generate another 160,000 multi-symmetry cases with randomly sampled coefficients. For these cases, the far-field amplitudes associated with the four symmetries are coherently summed within each circular polarization component before calculating the intensity profiles. The maximum relative amplitude of each anisotropic OAM component is reduced from 1 to 0.5, corresponding to a reduction in its maximum relative intensity from 1 to 0.25. This choice reflects the redistribution of harmonic energy among multiple OAM components.

The resulting dataset of 320,000 cases is split into \textit{training and validation sets} in an 80/20\% ratio. The training set is used to adjust the network weights by progressively minimizing the prediction error. The validation set is not used to update the weights but instead to assess generalization to unseen examples and guide the selection of the architecture and other hyperparameters. Additional 10,000 multi-symmetry cases are generated independently as a \textit{test set}, which is used neither for the training nor for model selection, providing a reliable evaluation of model's predictive performance. We use  \textit{tanh} as the \textit{activation function} \cite{iangoodfellowDeepLearning2016, casasromaDeepLearningPrincipios2020}, and the \textit{mean-squared error} (MSE) as the \textit{loss function}. The network is trained using the \textit{Adam} optimizer with a learning rate of $10^{-5}$. The phase of coefficients with amplitudes close to 0 have little influence on the far-field intensity patterns, potentially introducing large errors into the loss. For this reason, when evaluating the predictions we exclude the phase contribution to the MSE for coefficients whose amplitudes are below 10\% of the maximum.

\subsection{Solid-state high harmonic generation simulation models}

The solid-state HHG models used to validate the neural network predictions include two approaches: (1) a fully quantum description of the microscopic nonlinear response of a crystal, characterized by its band structure, and driven by an intense laser field; and (2), a macroscopic description of the response of the entire crystal layer, accounting for the spatial profile of driving beam, phase-matching effects, and  propagation of the generated harmonics to a far-field detector, approximating experimental conditions.

\subsubsection{Microscopic solid-state HHG response}

The nonlinear dynamics in the crystals considered in this work are calculated using the SBE \cite{SBE_1988, cistaro2023theoretical}, derived from the time-dependent Hamiltonian in the length-gauge $\mathcal{H}(t) = \mathcal{H}_0 - e \hat{\mathbf{r}} \cdot \mathbf{E}(t)$, where $e$ is the signed electron charge, $\hat{\mathbf{r}}$ is the position operator, $\mathcal{H}_0$ the field-free Hamiltonian and $\mathbf{E}(t)$ is the external electric field within the electric-dipole approximation. The total velocity of the charge carriers---i.e. the microscopic induced current---is computed as the trace of the velocity operator with the density matrix, integrated to the first Brillouin zone. Then the microscopic harmonic signal is obtained by Fourier transforming the time derivative of this current. The electronic structures of the materials considered in this work are calculated as follows.

On the one hand, for SLG, we use an effective two-band tight-binding model within the nearest-neighbour approximation, with a hopping energy of $2.97$~eV. Diagonalizing the tight-binding Hamiltonian $\mathcal{H}_0$ yields the band dispersions and the eigenstates used to construct the density matrix.

On the other hand, for AA- and AB-stacked BLG and monolayer hBN, the electronic structures are calculated using first-principles density-functional theory (DFT) as implemented in Quantum ESPRESSO \cite{giannozzi2009quantum, giannozzi2017advanced}. The relevant Kohn--Sham states are Wannierized using Wannier90 \cite{pizzi2020wannier90}, yielding a tight-binding basis comprising four $p_z$-like maximally localized Wannier functions per primitive unit cell for each bilayer (two per layer) and two for monolayer hBN (one centered on B and one on N). Within this basis, we obtain the real-space matrix elements of the field-free Hamiltonian $\mathcal{H}_0$, including onsite energies and hopping amplitudes, together with those of the position operator $\hat{\mathbf r}$. Wannier interpolation of these matrix elements yields the Hamiltonian and Berry-connection matrices in reciprocal space. These quantities enter the length-gauge SBE used for computing the microscopic harmonic signal for each material.

The linearly polarized driving laser field is modelled in the paraxial approximation as $\mathbf{E}(t) = E_0f(t)e^{-i2\pi c t/\lambda}\mathbf{e}_\alpha$, where $c$ is the speed of light, $\lambda$ the central wavelength, $E_0^2$ the peak intensity and $\mathbf{e}_\alpha$ is the unit polarization vector characterized by a tilt-angle $\alpha$. The polarization vector is contained within the crystal's lattice, perpendicular to the propagation direction. The temporal envelope is given by $f(t)=\sin ^2 (\pi t/\tau)$, where $\tau$ represents the total pulse duration.

\subsubsection{Macroscopic solid-state HHG propagation}

Starting from the microscopic nonlinear response, macroscopic HHG emission is computed using the framework developed for gas-phase HHG \cite{Carlos2010_macroscopic_HHG}, and previously applied to SLG to study transverse phase-matching effects \cite{BoyeroGarcia2021}, topological high-harmonic spectroscopy \cite{garcia-cabreraTopologicalHighharmonicSpectroscopy2024}, and the generation of solid-state attosecond pulse trains \cite{martíndomene2026attosecondpulsetrainsgraphene}. At the generation plane, the target layer is represented by randomly distributed elementary point sources within the discrete dipole approximation. Each source represents a cell large enough for an infinite-crystal description based on band theory to be valid, yet smaller than the driving wavelength, allowing the local phase and amplitude of the driving field to be considered homogeneous. The number of cells is large enough to ensure convergence of the results. The total electromagnetic field at the detector plane is computed as the coherent superposition of the far-field harmonic radiation emitted by these microscopic elementary sources by means of the Maxwell’s propagator.

For the macroscopic interaction, we consider LPVBs with two different radial profiles: (1) a standard Laguerre-Gauss (LG) profile, and (2), a flat-top profile. This comparison allows us to assess intensity-dependent effects in topological high-harmonic spectroscopy, as discussed in Sec. \ref{sec:macroHHG}. In both cases, the complex amplitudes of the driving beam's OAM components at the target plane are modelled using generalized LG profiles in cylindrical coordinates
\begin{equation}
    \mathrm{LG}^M_\ell(\rho,\phi) = \left\{ 1 - \left[ 1 - \left( \frac{\rho} {w_M} \right)^{|\ell|} e^{-\rho^2 / w_M^2} \right]^M \right\} e^{i \ell \phi},
\end{equation}
where $w_M$ is a beam waist parameter that depends on the flattening parameter $M$. The focused LPVBs at the target plane take the general form
\begin{multline}
    \mathrm{LPVB}^M_\ell(\rho,\phi) =\\ \frac{1}{\sqrt{2}} \left[ \mathrm{LG}^M_{-\ell}(\rho,\phi) \, \mathbf{e}_{\mathrm{LCP}} + \mathrm{LG}^M_{+\ell}(\rho,\phi) \, \mathbf{e}_{\mathrm{RCP}} \right],
\end{multline}
where $\mathbf{e}_{\mathrm{LCP}}$ and \(\mathbf{e}_{\mathrm{RCP}}\) are the circular polarization unit vectors. We consider LPVBs with $\mathcal{P}=1$, which allows us to set $\ell = 1$, and use $w_M \simeq w_0 /\sqrt{\ln M}$ for sufficiently large $M$, where $w_0$ is the beam waist of the standard LG profile ($M=1$), set to $w_0 = 30$~$\mu$m. For simulations with the flat-top radial profile, we use $M = 50$.

\section{Results and discussion} 

\subsection{Training and validation of the neural network}
\label{sec:MS}

\begin{figure}
\includegraphics[width=\columnwidth]{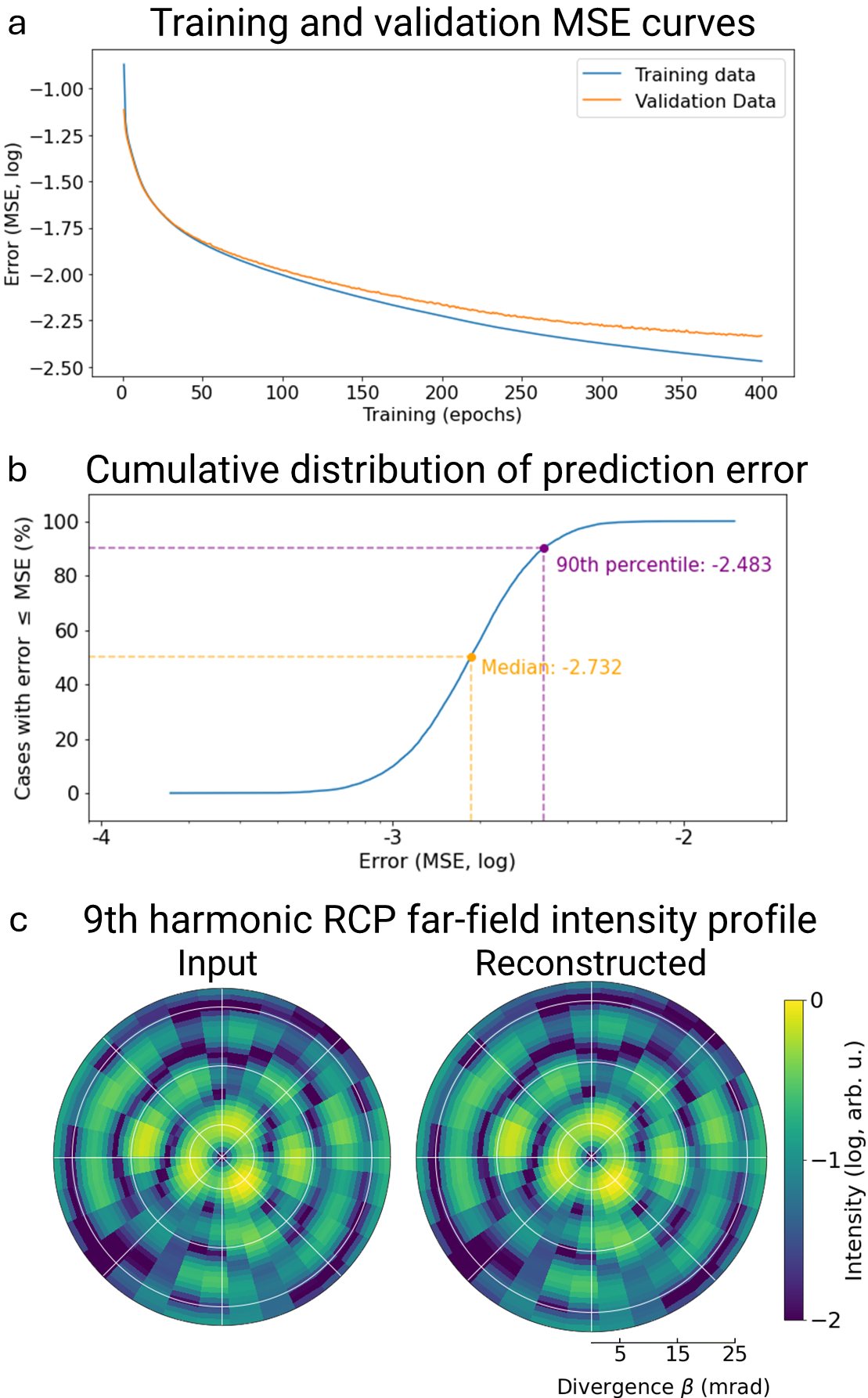}
\caption{Training, validation, and performance of the neural network. a) Evolution of $\log_{10}(\mathrm{MSE})$ for the training (blue) and validation (orange) sets over 400 epochs. The close agreement between the curves indicates no overfitting and an adequate generalization. b) Cumulative distribution of prediction errors for the multi-symmetry test set, with a median of $\log_{10}(\mathrm{MSE})=-2.732$ and a 90th percentile of $\log_{10}(\mathrm{MSE})=-2.483$. c) Input (left) and reconstructed (right) RCP far-field intensity profiles of the ninth harmonic for an example test case with an MSE close to the test-set median. The retrieved Fourier coefficients for this particular case are shown in Table \ref{tb:Multi}.}
\label{fig:MultiSymmetry}
\end{figure}

\begin{table}[!ht]
\caption{Comparison of the true and predicted amplitudes $a^{(N)}_{9,\pm1}$ and phases $\varphi^{(N)}_{9,\pm1}$ of the relative Fourier coefficients for the four rotational symmetries $N$ considered in the neural network. The results correspond to the test case shown in Fig. \ref{fig:MultiSymmetry}c, whose MSE is closest to the test-set median, $\log_{10}(\mathrm{MSE})=-2.732$. The agreement between the true and predicted coefficients demonstrates the accuracy of the neural network for this case.}
\resizebox{\columnwidth}{!}{%
\begin{ruledtabular}
\begin{tabular}{rrrrrrrrrr}
    $N$ & \multicolumn{2}{c}{$a^{(N)}_{9,1}$ [arb. u.]} & \multicolumn{2}{c}{$\varphi^{(N)}_{9,1}$ [rad]}
        & \multicolumn{2}{c}{$a^{(N)}_{9,-1}$ [arb. u.]}  & \multicolumn{2}{c}{$\varphi^{(N)}_{9,-1}$ [rad]}\\
        & True & Pred. & True & Pred. & True & Pred. & True & Pred. \\ \hline
2 & 0.316 & 0.286  & 0.474  & 0.532   & 0.112  & 0.104  & -1.139  & -1.127     \\
3 & 0.229 & 0.226  & -0.046 & -0.071  & 0.351  & 0.347  & -0.922  & -0.890     \\
4 & 0.070 & 0.055  &  0.567 & 0.471   & 0.321  & 0.325  & -0.134  & -0.080     \\
6 & 0.318 & 0.309  & 0.967  & 0.967   & 0.206  & 0.193  & 2.139   & 1.989      \\
\end{tabular}
\end{ruledtabular}
}
\label{tb:Multi}
\end{table}

Training and validation of our neural network are summarized in Fig. \ref{fig:MultiSymmetry}. We first represent in Fig. \ref{fig:MultiSymmetry}a the training and validation MSE curves over 400 epochs. Both curves decrease and remain close to each other, indicating no substantial overfitting and a successful generalization. Note that MSE is calculated from the network outputs, the normalized amplitudes and the sines and cosines of the phases of the Fourier coefficients. Fig. \ref{fig:MultiSymmetry}b shows the cumulative distribution of the MSE across the 10,000 independent test cases. The median $\log_{10}$(MSE) is --2.732 while the 90th percentile is --2.483, indicating that 90\% of the test cases have errors below this value.

Table \ref{tb:Multi} compares the true and predicted amplitudes $a_{9,s}^{(N)}$ and phases $\varphi_{9,s}^{(N)}$ of the two relative Fourier coefficients for each symmetry $N$, for a representative case with an MSE close to the median. For this case, we show in Fig. \ref{fig:MultiSymmetry}c the comparison between the input RCP far-field intensity of the 9-th harmonic (left) with the profile reconstructed from the predicted coefficients (right). The reconstructed far-field intensity profile closely reproduces the input pattern. The neural network is able to disentangle the contributions associated with twofold, threefold, and sixfold symmetries even when they have comparable weights and no single contribution clearly dominates the far-field intensity pattern. This artificial example represents a mixed-symmetry nonlinear response rather than that of a single ideal crystal with a well-defined rotational symmetry. Note that the neural network has been applied here---and through this work---to the information encoded in the 9th harmonic. Applying it to other harmonics would provide complementary information about the corresponding effective nonlinear susceptibilities.

\subsection{Application of AI-assisted topological high-harmonic spectroscopy to microscopic HHG simulations}

Once the neural network is trained and validated---establishing its ability to retrieve the Fourier coefficients from the corresponding far-field intensity patterns---we benchmark its predictions against HHG simulations. We first consider microscopic simulations in crystals exhibiting single-symmetry nonlinear response. We use graphene systems exhibiting a sixfold-symmetric harmonic response---SLG and BLG with AA and AB stacking. Then, we consider a crystal with mixed-symmetry responses, as monolayer hBN, which exhibits threefold and sixfold rotational symmetries. Finally we approach experimental conditions by comparing the neural network predictions with full macroscopic simulations in SLG.

\subsubsection{Single-symmetry responses}
\label{sec:Symmetry6}
 
The inverse problem in topological high-harmonic spectroscopy was solved analytically for SLG in \cite{garcia-cabreraTopologicalHighharmonicSpectroscopy2024}, which exhibits a sixfold rotational symmetry. Here we first demonstrate that AI-assisted topological high-harmonic spectroscopy successfully retrieves the Fourier coefficients from the harmonic intensity profiles of SLG.
 
We perform a series of microscopic HHG simulations in SLG for 60 orientations of the linearly polarized driving field, with polarization tilt angles $\alpha$ uniformly distributed over the interval $[0,\pi)$, thus covering all relative orientations between the electric field and the crystal. The driving field is modelled as a pulse with central wavelength of $\lambda=3$ $\mu$m, a duration of $\tau=8$ optical cycles ($29.1$ fs full width half maximum in intensity), and a peak intensity of $E_0^2=5\times10^{10}\,\mathrm{W\,cm^{-2}}$. We select the 9th harmonic order from the resulting HHG spectra. The selected 9th-harmonic fields are then arranged along a ring of radius $\rho_0=21.213$ $\mu$m, according to the ring of maximum intensity and local polarization orientation of a LPVB driving field with $\mathcal{P}=1$, to reconstruct the harmonic near field.
 
\begin{figure}
\includegraphics[width=\columnwidth]{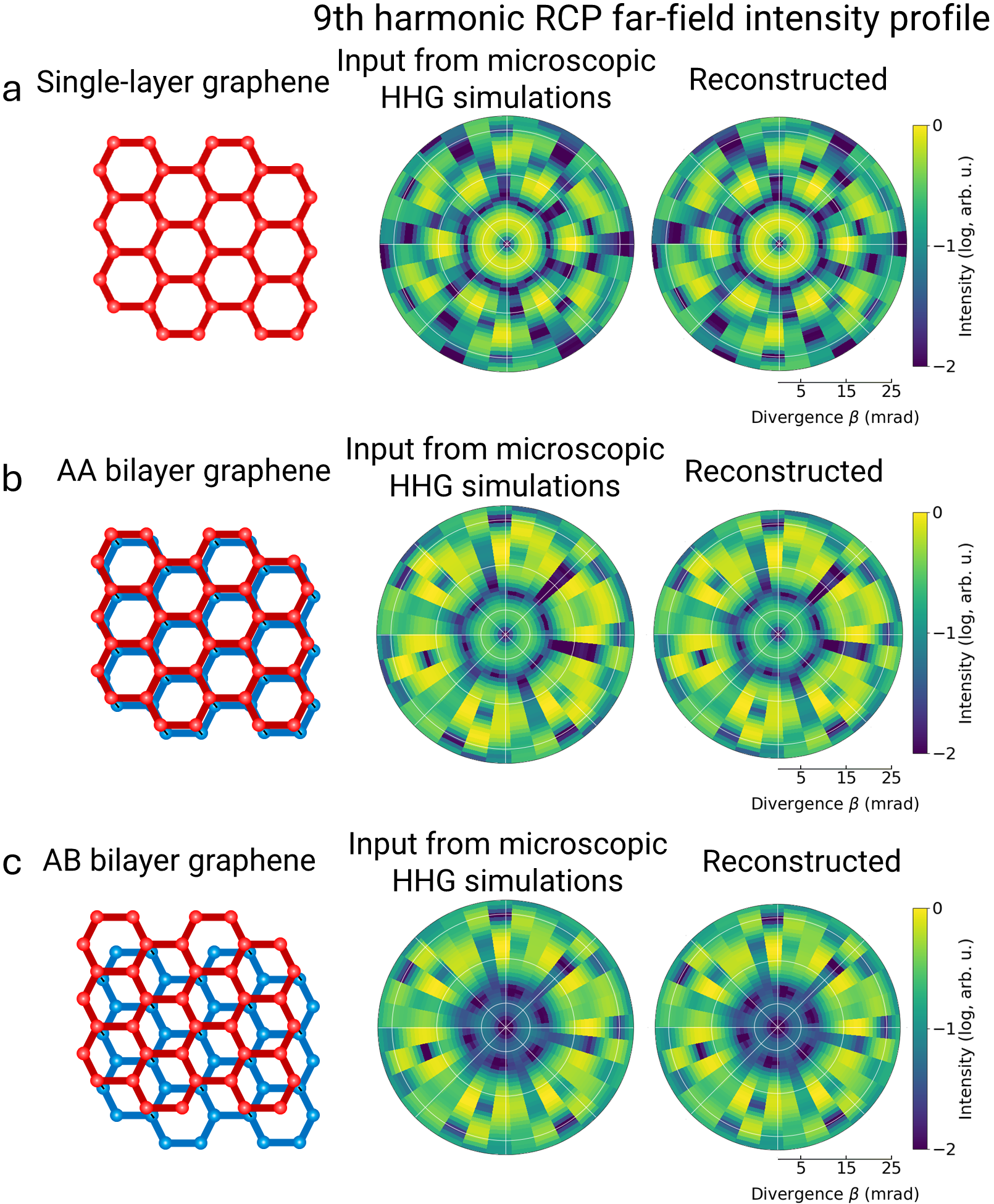}
\caption{
AI-assisted topological high-harmonic spectroscopy applied to microscopic HHG simulations in crystals with sixfold-symmetric responses ($N=6$): a) SLG, b) AA-stacked BLG, and c) AB-stacked BLG. Each row shows a schematic of the crystal lattice (left), the input RCP far-field intensity profile of the ninth harmonic obtained from microscopic HHG simulations (center), and the corresponding profile reconstructed from the predicted Fourier coefficients (right). The true and retrieved Fourier coefficients for each case are compared in Table \ref{tb:Symmetry6}.}
\label{fig:Symmetry6}
\end{figure}

We then follow a twofold strategy. First, we perform a Fourier transform along the azimuthal coordinate at $\rho_0$ to extract the Fourier coefficients from the numerical simulations, listed in the first row of Table \ref{tb:Symmetry6} and labelled as ``True".
Note that consistent with the sixfold symmetry of the graphene lattice, the extracted---and then predicted---amplitudes of coefficients associated with $N\neq6$ fall below $10^{-3}$. We therefore report only the coefficients for $N=6$.
Second, we numerically evaluate the Fraunhofer integral of the 9th harmonic near-field to obtain the far-field RCP and LCP intensity profiles, which serve as inputs for the neural network. The resulting RCP far-field intensity profile is shown in the left panel of Fig. \ref{fig:Symmetry6}a.

\begin{table}[!ht]
\caption{Comparison of the true and predicted amplitudes $a^{(6)}_{9,\pm1}$ and phases $\varphi^{(6)}_{9,\pm1}$ of the relative Fourier coefficients for the simulated materials with a sixfold-symmetric response ($N=6$): SLG  with $\log_{10}(\mathrm{MSE})=-2.542$, AA-stacked BLG with $\log_{10}(\mathrm{MSE})=-2.838$, and AB-stacked BLG with $\log_{10}(\mathrm{MSE})=-2.722$. The results correspond to the cases shown in Fig. \ref{fig:Symmetry6}. Components with $N\neq6$, whose predicted amplitudes are small, are omitted from the table. The agreement between the true and predicted coefficients demonstrates the accuracy of the network for these single-symmetry responses, benchmarked against microscopic HHG simulations.}
\resizebox{\columnwidth}{!}{%
\begin{ruledtabular}
\begin{tabular}{rrrrrrrrrr}
    Material & \multicolumn{2}{c}{$a^{(6)}_{9,1}$ [arb. u.]} & \multicolumn{2}{c}{$\varphi^{(6)}_{9,1}$ [rad]}
        & \multicolumn{2}{c}{$a^{(6)}_{9,-1}$ [arb. u.]}  & \multicolumn{2}{c}{$\varphi^{(6)}_{9,-1}$ [rad]}\\
        & True & Pred. & True & Pred. & True & Pred. & True & Pred. \\ \hline
SLG    & 0.777 & 0.768  & 0.716  & 0.878   & 0.824  & 0.817  & -2.803  & -2.871     \\
AA-BLG & 1.991 & 2.035  & 1.341  & 1.323  & 1.562  & 1.796  & -1.373  & -1.349     \\
AB-BLG & 4.448 & 4.394  &  1.527 & 1.555   & 4.106  & 4.310  & -1.239  & -1.201    \\
\end{tabular}
\end{ruledtabular}
}
\label{tb:Symmetry6}
\end{table}

The Fourier coefficients predicted by the neural network for SLG are listed in the first row of Table \ref{tb:Symmetry6}, whereas the reconstructed far-field intensity profile of the 9th harmonic is shown in the right panel of Fig. \ref{fig:Symmetry6}a. The resulting error,  $\log_{10}(\textrm{MSE})=-2.542$, lies below the 90th percentile  of the test-set error distribution. Thus, the neural network trained on multiple symmetries accurately retrieves the non-linear response of a crystal with well-defined rotational symmetry---such as SLG---demonstrating that AI-assisted harmonic topological spectroscopic reproduces the analytical results reported in \cite{garcia-cabreraTopologicalHighharmonicSpectroscopy2024}.

To assess the generality of AI-assisted topological high-harmonic spectroscopy, we consider additional crystals with sixfold rotational symmetry but different nonlinear susceptibilities. Specifically, we consider BLG in AA and AB stacking configurations. The microscopic HHG simulations reveal larger relative Fourier amplitudes, requiring an extension of their range to $a_{9,s}^{(N)}\in[0,5]$. For the updated network, the median and 90th percentile of the test-set $\log$(MSE) distribution are --2.024 and --1.756, respectively.

The Fourier coefficients predicted by the neural network for AA-BLG and AB-BLG are listed in the second and third rows of Table \ref{tb:Symmetry6}, respectively, whereas the reconstructed far-field intensity profiles of the 9th harmonic are shown in the right panels of Figs. \ref{fig:Symmetry6}b and \ref{fig:Symmetry6}c. As in the case of SLG, we report only the coefficients for $N=6$.
The resulting errors are $\log_{10}(\textrm{MSE})=-2.838$ and $\log_{10}(\textrm{MSE})=-2.722$ for AA-BLG and AB-BLG, respectively, both below the median of the test-set error distribution. These results demonstrate that the neural network distinguishes and accurately retrieves different nonlinear responses within the same rotational symmetry from their characteristic far-field harmonic intensity profiles.

\subsubsection{Mixed-symmetry responses}

We further extend our analysis to nonlinear responses containing contributions associated with multiple rotational symmetries. Specifically, we apply the neural network to HHG in monolayer hBN, whose response contains both threefold and sixfold angular components, the latter corresponding to the second Fourier harmonic of the threefold dependence. Higher-order threefold Fourier components are also present in the HHG simulations, but their contributions are negligible within the far-field angular range considered. As indicated by Eq. \ref{eq:farf}, the associated higher-order OAM modes primarily contribute at larger divergence angles.

\begin{table}[!ht]
\caption{Comparison of the true and predicted amplitudes $a^{(N)}_{9,\pm1}$ and phases $\varphi^{(N)}_{9,\pm1}$ of the relative Fourier coefficients for monolayer hBN, whose response is dominated by threefold and sixfold components. The overall error is $\log_{10}(\mathrm{MSE})=-1.029$, and the corresponding far-field intensity profiles are shown in Fig. \ref{fig:hBN}. The omission of higher-order Fourier coefficients from the model can increase the overall error by introducing spurious contributions, such as the $N=4$ component predicted here. Nevertheless, the dominant coefficients show good agreement with those extracted from microscopic HHG simulations.}
\resizebox{\columnwidth}{!}{%
\begin{ruledtabular}
\begin{tabular}{rrrrrrrrrr}
    $N$ & \multicolumn{2}{c}{$a^{(N)}_{9,1}$ [arb. u.]} & \multicolumn{2}{c}{$\varphi^{(N)}_{9,1}$ [rad]}
        & \multicolumn{2}{c}{$a^{(N)}_{9,-1}$ [arb. u.]}  & \multicolumn{2}{c}{$\varphi^{(N)}_{9,-1}$ [rad]}\\
        & True & Pred. & True & Pred. & True & Pred. & True & Pred. \\ \hline
2 & 0.014 & 0.109  & 3.115  & -1.184   & 0.011  & 0.057  & 2.966  & -2.106     \\
3 & 0.963 & 0.977  & 1.691  & 1.793  & 0.929  & 0.926  & 2.211  & 2.367     \\
4 & 0.010 & 0.421  &  0.251 & -1.096   & 0.015  & 0.387  & -2.463  & 2.831    \\
6 & 0.570 & 0.710  &  -2.234 & -2.267   & 0.737  & 0.848  & -1.999  & -1.982    \\
\end{tabular}
\end{ruledtabular}
}
\label{tb:hBN}
\end{table}

\begin{figure}
\includegraphics[width=\columnwidth]{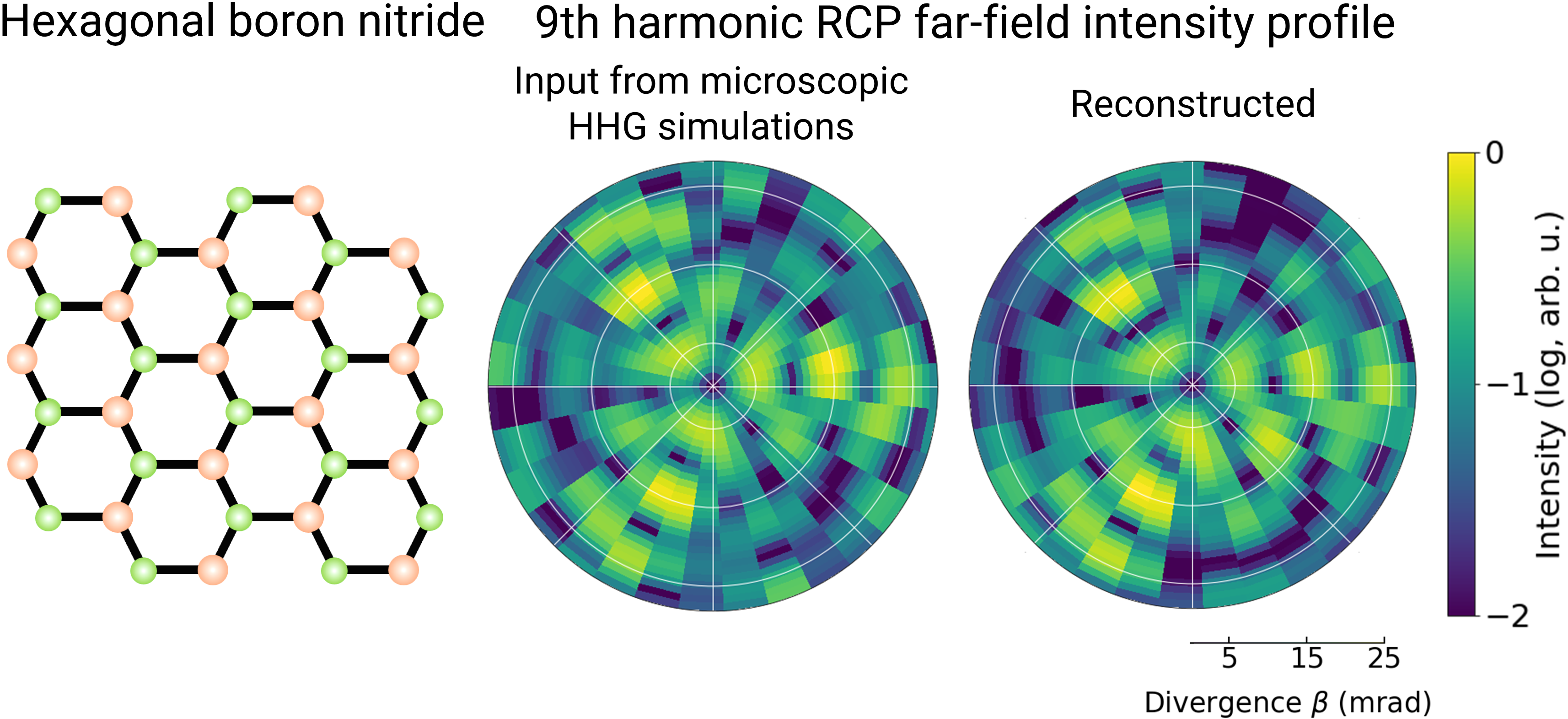}
\caption{AI-assisted topological high-harmonic spectroscopy applied to microscopic HHG simulations in monolayer hBN, whose response contains both threefold ($N=3$) and sixfold ($N=6$) components. The panels show a schematic of the crystal lattice (left), the input RCP far-field intensity profile of the 9-th harmonic obtained from microscopic HHG simulations (center), and the corresponding profile reconstructed from the predicted Fourier coefficients (right). The true and retrieved Fourier coefficients for each symmetry are compared in Table \ref{tb:hBN}.}
\label{fig:hBN}
\end{figure}

The Fourier coefficients predicted by the neural network for hBN for $N=2,3,4,6$  are listed in Table \ref{tb:hBN}, and the input and reconstructed far-field intensity profiles of the 9th harmonic are shown in Fig. \ref{fig:hBN}. The overall $\log_{10}$(MSE) error is --1.029, while the corresponding ones for the main rotational symmetries, $N=3$ and $N=6$, are --2.222 and --1.509, respectively. These results demonstrate that the neural network retrieves the dominant threefold and sixfold contributions to the nonlinear response. However, the omission of higher-order Fourier coefficients from the model can increase the overall error by introducing spurious contributions, such as the $N=4$ component predicted here. Extending the network to include higher-order coefficients could reduce these artifacts, at the expense of increased memory requirements and larger datasets, which is beyond the scope of this work. This example thus illustrates both the potential of the method and the challenges of retrieving mixed-symmetry responses.

\subsection{Application of AI-assisted topological high-harmonic spectroscopy to macroscopic HHG simulations}
\label{sec:macroHHG}

Applying AI-assisted topological high-harmonic spectroscopy under realistic experimental conditions requires accounting for harmonic generation across the full spatial profile of the driving LPVB, rather than only on an infinitesimally thin ring of radius $\rho_0$, as done up to now in this work. In this section, we describe how the neural network training is adapted to account for this spatial dependence and test its predictions against macroscopic HHG simulations in SLG driven by LPVBs with flat-top and Gaussian radial profiles.

To generate the dataset of our network, we start from the Fourier coefficients associated to the $q$-th order nonlinear response, but now we model the near-field harmonic profile as a sum of infinitely thin rings with different radii and identical Fourier coefficients. Following the integration procedure explained in Sec. \ref{sec:FarFieldModel}, the far-field harmonic profile takes the form of a superposition of Bessel functions. The RCP and LCP amplitudes of the $q$th-order harmonic, $U_q^{\pm}(\beta,\theta)$, are thus given by,
\begin{equation}
\begin{split}
U_q^{\pm}(\beta,\theta)&=-i\kappa\frac{e^{i\kappa D}}{D}\sum_j \rho_j\Delta\rho  \sum_{s=-1}^1 c_{q,s}^{\pm}(\rho_0) \\
& \times e^{\mp i(Ns+1)\mathcal{P}\pi/2} J_{\pm (Ns+1)\mathcal{P}}\left(\kappa \rho_j \beta \right)e^{\pm i (Ns+1)\mathcal{P}\theta},
\label{eq:MBessel}
\end{split}
\end{equation}
where $\rho_j$ samples the interval $[\rho_0-\delta/2,\rho_0+\delta/2]$, $\Delta\rho$ is the radial sampling step, and  $\delta$ is the total width of the harmonic near-field ring. Note that allowing the Fourier coefficients to vary independently with radius would provide a more accurate description, but would substantially increase the number of network outputs and the training data required, making this extension computationally impractical.

\begin{figure}
\includegraphics[width=\columnwidth]{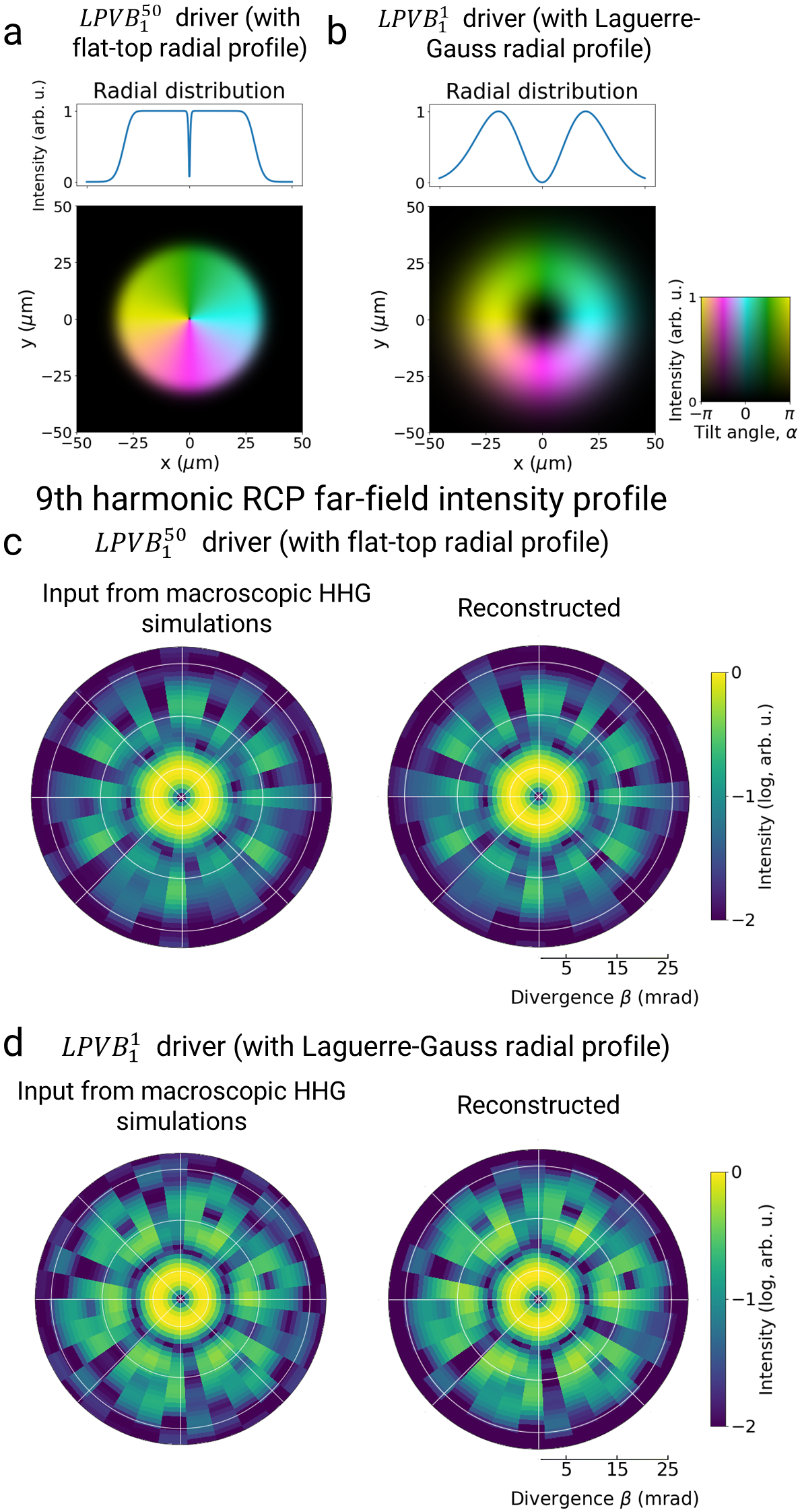}
\caption{AI-assisted topological high-harmonic spectroscopy applied to macroscopic HHG simulations in SLG driven by two beams: a) a $\mathrm{LPVB}_1^{50}$ with a flat-top radial profile, and b) a $\mathrm{LPVB}_1^{1}$ with a Laguerre-Gaussian radial profile. Panels c) and d) show the input RCP far-field intensity profiles of the ninth harmonic obtained from macroscopic HHG simulations (left) and the corresponding profiles reconstructed from the predicted Fourier coefficients (right), for the $\mathrm{LPVB}_1^{50}$ and $\mathrm{LPVB}_1^{1}$ drivers, respectively. The true and retrieved Fourier coefficients for the $N=6$ response of SLG obtained for each driving beam are compared in Table \ref{tb:SergioMacro}.}
\label{fig:SergioMacro}
\end{figure}

\begin{table}[!ht]
\caption{Comparison of the true amplitudes $a^{(6)}_{9,\pm1}$ and phases $\varphi^{(6)}_{9,\pm1}$ of the relative Fourier coefficients, extracted from microscopic HHG simulations, with those predicted by the neural network from macroscopic HHG far-field intensity profiles in SLG ($N=6$). Two driving beams are considered: a $\mathrm{LPVB}_1^{50}$ with a flat-top radial profile, yielding $\log_{10}(\mathrm{MSE})=-1.671$, and a $\mathrm{LPVB}_1^{1}$ with a Laguerre-Gaussian radial profile, yielding $\log_{10}(\mathrm{MSE})=-1.081$. The results correspond to the cases shown in Fig. \ref{fig:SergioMacro}. Components with $N\neq6$, whose predicted amplitudes are small, are omitted from the table. The lower MSE for the flat-top case highlights the sensitivity of the retrieval to the driving beam's radial intensity distribution. For the Laguerre-Gaussian driver, the retrieved coefficients are interpreted as effective parameters reflecting both the target's nonlinear response and the coherent superposition of harmonic emission across the layer.}
\resizebox{\columnwidth}{!}{%
\begin{ruledtabular}
\begin{tabular}{rrrrrrrrrr}
    Driver & \multicolumn{2}{c}{$a^{(6)}_{9,1}$ [arb. u.]} & \multicolumn{2}{c}{$\varphi^{(6)}_{9,1}$ [rad]}
        & \multicolumn{2}{c}{$a^{(6)}_{9,-1}$ [arb. u.]}  & \multicolumn{2}{c}{$\varphi^{(6)}_{9,-1}$ [rad]}\\
        & True & Pred. & True & Pred. & True & Pred. & True & Pred. \\ \hline
LPVB$_1^{50}$  & 0.777 & 0.653  & 0.716  & 0.909   & 0.824  & 0.642  & -2.803  & -2.671    \\
LPVB$_1^{1}$ & 0.777 & 0.695  & 0.716  & 1.643  & 0.824  & 0.766  & -2.803  & -2.420     \\
\end{tabular}
\end{ruledtabular}
}
\label{tb:SergioMacro}
\end{table}

First, we study macroscopic HHG driven by a $\mathrm{LPVB}_1^{50}$ beam, i.e. a LPVB with $\mathcal{P}=1$ and a flat-top radial profile, as depicted in Fig. \ref{fig:SergioMacro}a. Color indicates the polarization tilt-angle $\alpha$, while brightness represents intensity. The inset shows the radial intensity profile. This flat-top profile is chosen to minimize radially-spatial variations in the intensity-dependent phase of the HHG emission. A similar strategy  has been used in gas-phase HHG to reveal microscopic HHG features that would otherwise be washed out by macroscopic propagation and intensity-dependent phase-matching effects \cite{Dubrouil2011,Boutu2011,Hernandez2016,Veyrinas_2023}. For this case, we have trained and validated our neural network using the model in Eq. \ref{eq:MBessel}, with $\rho_0=w_M=15.17$ $\mu$m and $\delta=$ 24 $\mu$m. These values were chosen to approximate as closely as possible the spatial characteristics of the $\mathrm{LPVB}_1^{50}$ driver considered here. The median and 90th percentile of the test-set $\log_{10}$(MSE) distribution are --2.375 and --2.129, respectively.

The RCP far-field intensity profile of the 9th harmonic obtained from macroscopic HHG simulations in SLG---and used as input to our neural network---is shown in the left panel of Fig. \ref{fig:SergioMacro}c, whereas the reconstructed profile is shown in the corresponding right panel. The Fourier coefficients predicted by the neural network are listed in the first row of Table \ref{tb:SergioMacro}. As in the case of the microscopic HHG case, we report only the coefficients for $N=6$, as the extracted---and then predicted---amplitudes of coefficients associated with $N\neq6$ fall below $10^{-2}$. The resulting error is $\log_{10}(\mathrm{MSE})=-1.671$. Although the neural network is trained on spatial distributions with radially uniform Fourier coefficients, it retrieves the coefficients from full macroscopic HHG simulations with reasonable accuracy. Thus, these results support the applicability of AI-assisted topological high-harmonic spectroscopy under macroscopic conditions, even when the Fourier coefficients are assumed to be independent of the radial coordinate.

Second, to assess intensity-dependent effects of the driving beam in our model, we study macroscopic HHG driven by a $\mathrm{LPVB}_1^{1}$ beam, i.e. a LPVB with $\mathcal{P}=1$ and a Laguerre-Gaussian radial profile, as depicted in Fig. \ref{fig:SergioMacro}b. For this case, we trained and validated our neural network using the model in Eq. \ref{eq:MBessel}, with $\rho_0=21.213$ $\mu$m and $\delta=$ 12 $\mu$m, to approximate as closely as possible the spatial characteristics of the $\mathrm{LPVB}_1^{1}$ driver. The median and 90th percentile of the test-set $\log_{10}$(MSE) distribution are --2.538 and --2.287, respectively.

The reconstructed RCP far-field intensity profile of the 9th harmonic is shown in Fig. \ref{fig:SergioMacro}d. The predicted Fourier coefficients---listed in the second row of Table \ref{tb:SergioMacro}---yield an error of $\log_{10}(\mathrm{MSE})=-1.081$. The higher MSE compared with the flat-top case indicates that spatial variations in the driving intensity affect the retrieval, highlighting the sensitivity of AI-assisted topological high-harmonic spectroscopy to the beam's radial intensity distribution. The retrieved coefficients should therefore be interpreted as effective parameters reflecting both the target's nonlinear response and the coherent superposition of harmonic emission across the illuminated region. As discussed above, allowing the Fourier coefficients to vary independently with radius would provide a more accurate description, at the cost of increased memory and computational requirements, which is beyond the scope of this work.

\section{Conclusions} 

We have demonstrated that AI-assisted topological high-harmonic spectroscopy enables the quantitative retrieval of the anisotropic nonlinear response of thin crystalline materials from the far-field intensity profiles of their harmonic emission. We have developed, trained, and validated a convolutional neural network that recovers the complex Fourier coefficients describing the nonlinear material response from the RCP/LCP far-field harmonic intensity distributions. The application of this method does not require phase-resolved measurements of the EUV radiation, which are experimentally challenging.

The neural network not only retrieves the nonlinear response but also distinguishes contributions associated with different rotational symmetries. Comparisons with microscopic HHG simulations for single-layer graphene and AA- and AB-stacked bilayer graphene confirms the ability of the method to distinguish between different nonlinear responses in materials sharing the same rotational symmetry. Moreover, its application to monolayer hBN further demonstrates that it can resolve multiple angular components within a single material. Comparison with macroscopic HHG simulations supports the applicability of the method under the macroscopic conditions considered. 

However, intensity-dependent effects in the harmonic emission mean that the retrieved coefficients encode information about both the material response and the coherent emission across the illuminated region. A deeper understanding of intensity-dependent HHG in solids could help relate these effective coefficients to the local nonlinear response, extending the quantitative scope of the technique.

These results provide a basis for all-optical characterization of anisotropic nonlinear responses through harmonic intensity measurements, with the topology of light encoding information about the material. The ability to separate angular contributions also motivates future applications to more complex systems, including polycrystals and van der Waals heterostructures, where the relation between the measured response and the constituent structures remains to be explored. Further developments could enable probing exciton dynamics, or the detection of lattice defects, opening possibilities for optical quality control in the scalable fabrication of atomically-thin layered materials.

\section*{Acknowledgements} 

This project has received funding from the European Research Council (ERC) under the European Union’s Horizon Europe research and innovation programme (ERC Proof of Concept grant agreement No. 101247206). We acknowledge funding from Ministerio de Ciencia e Innovaci\'on (Grant PID2025-173282NB-I00) and the Department of Education of the Junta de Castilla y Le\'on and FEDER Funds (No. SA108P24 and Escalera de Excelencia CLU-2023-1-02). D.H.-S. acknowledges financial support from the University of Salamanca through its Predoctoral Fellowship Programme III (2024), co-funded by Banco Santander.

\section*{Data availability}
The data supporting the findings of this study are available from the corresponding author upon reasonable request.

\section*{Conflict of interest}
The authors declare no conflicts of interest.

\bibliographystyle{ieeetr}
\bibliography{bibliography}

\end{document}